\documentclass[aps,superscriptaddress,nofootinbib,longbibliography,showkeys]{revtex4-2}
\usepackage{amssymb}
\usepackage{amsmath}
\usepackage{txfonts}
\usepackage{epsfig}
\usepackage{caption}
\usepackage{float}
\usepackage{indentfirst}
\usepackage[breaklinks,colorlinks,citecolor=green,urlcolor=magenta,linkcolor=red]{hyperref}
\usepackage{caption}
\usepackage{graphicx,subfig}
\usepackage{epstopdf}
\usepackage{diagbox}
\usepackage{threeparttable}
\usepackage{array}
\usepackage{multirow}

\usepackage[markup=default]{changes}
\newcommand{\stkout}[1]{\ifmmode\text{\sout{\ensuremath{#1}}}\else\sout{#1}\fi}
\setdeletedmarkup{\stkout{#1}}

\usepackage{CJKutf8}

\begin{document}
	

\title{The forward-backward asymmetry induced $CP$ asymmetry in ${\overline{B}}^{0}\rightarrow K^{-}\pi^{+}\pi^{0}$ in phase space around the resonances ${\overline{K}}^{*}(892)^{0}$ and ${\overline{K}}^{*}_{0}(700)$}

\author{Jian-Yu Yang}
\email{yangjy@stu.usc.edu.cn}
\affiliation{School of Nuclear Science and Technology, University of South China, Hengyang, Hunan 421001, China.}

\author{Yu-Jie Zhao} 
\email{zhaoyj\_usc@163.com}
\affiliation{School of Nuclear Science and Technology, University of South China, Hengyang, Hunan 421001, China.}

\author{Jing-Juan Qi} 
\email{qijj@mail.bnu.edu.cn}
\affiliation{College of Information and Intelligence Engineering, Zhejiang Wanli University, Ningbo, Zhejiang 315101, China}

\author{Zhen-Hua Zhang} 
\email{corresponding author, zhangzh@usc.edu.cn}
\affiliation{School of Nuclear Science and Technology, University of South China, Hengyang, Hunan 421001, China.}
\date{\today}

\begin{abstract}
The interference between amplitudes corresponding to different intermediate resonances plays an important role in generating large CP asymmetries in phase space in multi-body decays of bottom and charmed mesons.
In this paper, we study the CP violation in the decay channel ${\overline{B}}^{0}\rightarrow K^{-}\pi^{+}\pi^{0}$ in phase space region where the intermediate resonances $\overline{K}^{*}(892)^{0}$ and ${\overline{K}^{*}_{0}(700)}$ dominate.
The Forward-Backward Asymmetry (FBA) and the CP asymmetry induced by FBA (FB-CPA), which are closely related to the interference effects between the two aforementioned resonances, are especially investigated.
The non-trivial correlation between FBA and FB-CPA is analyzed.
The analysis indicates that the FB-CPAs around the resonance $\overline{K}^{*}(892)^{0}$ can be as large as about 35\%, which can be potentially accessible by Belle and Belle-II collaborations in the near future.
\end{abstract}

\maketitle


\section{\label{sc:introduction} Introduction}
CP violation (CPV) has played an important role in particle physics since its first discovery in the decay of $K^{0}_{L}$ mesons in 1964 \cite{Christenson:1964fg}. 
Besides the strange meson systems, CP violation has been observed in $B$ meson, $B_s$ meson, $D$ meson systems \cite{BaBar:2001pki,Belle:2001zzw,LHCb:2013syl,LHCb:2019hro}, and recently, in the $\Lambda_b^0$ decay process $\Lambda_b^0\to p K^- \pi^+\pi^-$ \cite{LHCb:2025ray}, all of which are consistent with the description of the the Standard Model (SM) of particle physics,
within which CPV is described by  an irreducible complex phase in the Cabibbo-Kobayashi-Maskawa (CKM) quark mixing matrix \cite{Cabibbo:1963yz,Kobayashi:1973fv}. 
According to the Sakharov's criteria \cite{Sakharov:1967dj}, CPV is also an important necessary condition to explain the Baryon Asymmetry of the Universe (BAU).
However, it is widely accepted that CPV within the SM is by far not enough to explain the current BAU \cite{Shaposhnikov:1986jp,Shaposhnikov:1987tw,Shaposhnikov:1987pf,Farrar:1993sp,Huet:1994jb}.
Hence a better understanding of CPV is of great importance both to particle physics and to Cosmology.

Thanks to the increasing amount of data, CPV is now able to be investigated in some multi-body decays of bottom and charmed hadrons, in which CP asymmetries (CPAs) with great detailed structures were observed in different regions of phase space \cite{LHCb:2013ptu,LHCb:2013lcl,LHCb:2014mir,LHCb:2019jta,LHCb:2022fpg,LHCb:2024rkp}.
Some of the regional CPAs in charged $B$ meson multi-body decays can be as large as 80\%, which represent the largest CPAs ever observed in laboratories.
In contrast, the overall CPAs of these decays are considerably smaller. 

The structures of regional CPAs in phase space in three-body decay of charged $B$ mesons indicates that the interference between different intermediate resonances play an important role in generating the aforementioned large regional CPAs.
Take the decay $B^\pm\to\pi^\pm\pi^+\pi^-$ as an example.
The results of the LHCb collaboration showed that large regional CPA is located around the intermediate resonances $\rho(770)^0$.
The behaviour of the large regional CPA can be understood as an consequence of the interference effect between the $\rho(770)^0$ resonance and some $s$-wave structure \cite{Zhang:2013oqa,Cheng:2022ysn}. 
Since $\rho(770)^0$ has spin-1, meaning that the $\pi^+\pi^-$ system originating from $\rho(770)^0$ with the residence pion are in $p$-wave, hence a Forward-Backward Asymmetry (FBA) in the distributions of the final particles is produced because of the aforementioned interfering effects.
Consequently, the CPV corresponding to this interfering effect, which is the dominant contributions to the aforementioned large regional CPAs, can also result in CPA corresponding to the FBA (FB-CPA) \cite{Zhang:2021zhr,Wei:2022zuf}.

In this paper, we study a similar interfering effect, which is the interference of amplitudes corresponding to $\overline{K}^\ast(892)^0$ and $\overline{K}_0^\ast(700)$ in the decay channel of ${\overline{B}}^{0}\rightarrow K^{-}\pi^{+}\pi^{0}$. 
This decay channel has been studied by the BaBar and the Belle collaborations, no CPV is established, mainly because of the low statistics \cite{Belle:2004khm,BaBar:2004bcz,BaBar:2007hmp,BaBar:2011vfx}.
The analysis of this paper will show that the interference of $\overline{K}^\ast(892)^0$ and $\overline{K}^\ast_0(700)$ may result in CPV which has the potential to be detectable with more data through the measurements of the observables such as FB-CPA. 
There is very strong evidence that the interference of $\overline{K}^\ast(892)^0$ and $\overline{K}^\ast_0(700)$ in the baryon-anti-baryon-production decay channel $\overline{B}^0\to p\bar{p} K^- \pi^+$ can lead to large CP asymmetries \cite{Zhang:2025mne}.
Moreover, the contribution of the interference of $\overline{K}^\ast(892)^0$ and $\overline{K}^\ast_0(700)$ in the $\Lambda_b^0$ decay process $\Lambda_b^0\to p K^- \pi^+\pi^-$ \cite{LHCb:2025ray} may also potentially related to the observed regional CPAs $(5.3\pm1.3\pm0.2)\%$, which deserves further investigation.

The remainder of this paper is organized as follows. In Sec. \ref{sec:FBA}, we present a brief introduction of FBA and FB-CPA. 
In Sec. \ref{sec:DecayAmp}, we present the decay amplitudes
of the decay channel ${{\overline{B}}^{0}\rightarrow K^{-}\pi^{+}\pi^{0}}$, where we focus mainly on the phase space region where both the intermediate resonances $\overline{K}^{*}(892)^{0}$ and $\overline{K}^{*}_{0}(700)$ dominate. 
In Sec. \ref{sec:NumDis}, we present the numerical results and discussion. 
The last section (Sec. \ref{sec:SumCon}) is our summary and conclusions.


\section{\label{sec:FBA}\boldmath The forward-backward asymmetry induced $CP$ asymmetry}
\begin{figure}[H]
  \centering
  \includegraphics[width=0.5\textwidth]{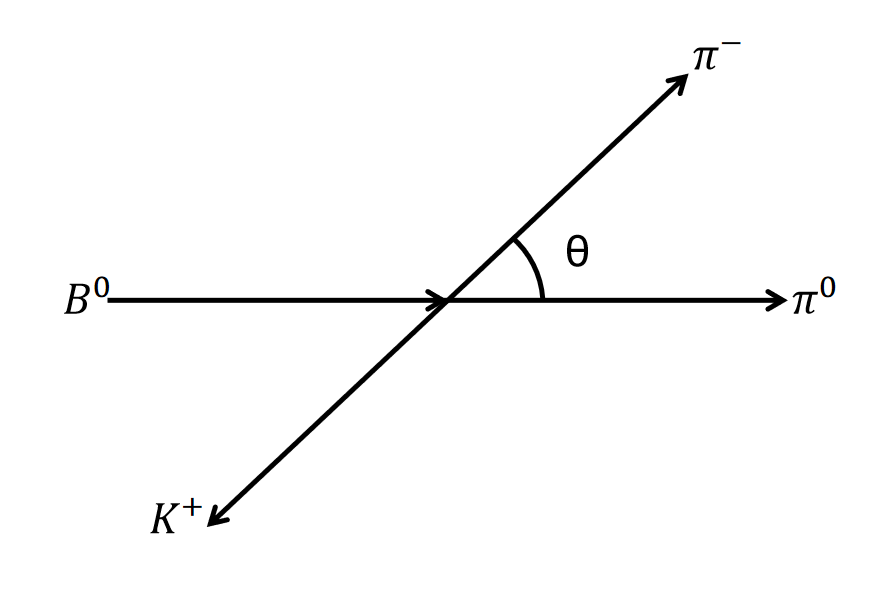}\\
  \caption{The definition of $\theta$ in the $B^{0}\rightarrow K^{+}\pi^{-}\pi^{0}$ decay channel.}
  \label{FIG:1}
\end{figure}

For a three-body decay process $H\to M_1 M_2 M_3$, with $H$ being a heavy pseudo-scalar meson, and $M_1,M_2,M_3$ being three light pseudo-scalar mesons, we can parameterize the decay amplitude as
\begin{equation}
\begin{aligned}
\mathcal{M}=\sum\limits_{l}\mathcal{A}_{l}P_{l}(\cos\theta),
\label{eq11}
\end{aligned}
\end{equation}
where $P_{l}$ is the Legendre polynomial of order $l$, with $l$ being the quantum number of the angular momentum between $M_3$ and the $M_1M_2$ system,
$\theta$ is the helicity angle which is defined as the angle between the momentum of $M_1$ and that of $H$ (or $M_3$) in the centre-of-mass (C.O.M.) frame of the $M_1M_2$ system, 
$\mathcal{A}_l$ is the amplitude of the $l$-th wave.
In Fig. \ref{FIG:1} we present a illustration of the helicity angle $\theta$ for the case $B^0\to K^+\pi^-\pi^0$, as this is just the decay channel which we will focus on.
Note that $l$ is also the spin quantum number of the $M_1M_2$ system.

The forward-backward asymmetry (FBA) in the decay channel $H\to M_1 M_2 M_3$ can be defined as the difference between the event yields for the $M_3$ flying forward ($\cos\theta>0$) and backward ($\cos\theta<0$) in the C.O.M. frame of the $M_1M_2$ systerm, which can be expressed as 
\begin{equation}
 A^{FB}=\frac{\int_0^1\langle\left|\mathcal{M}\right|^2\rangle d\cos\theta-\int_{-1}^0\langle\left|\mathcal{M}\right|^2\rangle d\cos\theta} {\int_{-1}^1\langle \left|\mathcal{M}\right|^2\rangle d\cos\theta},
\label{eq12}
\end{equation}
where the notion ``$\langle \cdots\rangle$'' represents the integration over phase space other than $\cos\theta$.
It can be straightforwardly shown that \cite{Wei:2022zuf}
\begin{equation}\label{eq:AFB}
  A^{FB}= \frac{2}{\sum_{j} \left[{\langle \left|a_j\right|^2\rangle}/{(2j+1)}\right]}\sum_{{\text{even~}} l \atop {\text{odd~}} k} f_{lk}\Re\left(\langle \mathcal{A}_{l}\mathcal{A}_{k}^{\ast}\rangle\right),
\end{equation}
where $f_{lk}\equiv\int_0^1 P_{l}P_{k}d\cos\theta=\frac{(-)^{(l+k+1)/2}l!k!}{2^{l+k-1}(l-k)(l+k+1)[(l/2)!]^2\{[(k-1)/2]!\}^2}$.
From the above equation one can clearly see that FBA is caused by the real part of the interference between even- and odd-wave amplitudes $ \Re( \langle a_{l}a_{k}^{\ast}\rangle)$.
In the simplest case, the amplitudes is dominated by the $S$- and $P$-wave amplitudes ($l=0$, and 1, respectively), so that 
the FBA can be expressed as
\begin{equation}
  A^{FB}= \frac{\Re(\langle \mathcal{A}_P \mathcal{A}_S^\ast\rangle)}{\left\langle\left| \mathcal{A}_P\right|^2\right\rangle/3+\left\langle\left| \mathcal{A}_S\right|^2\right\rangle}.
\end{equation}
The FBA include CP asymmetry (FB-CPA) can be defined as
\footnote{In Ref. \cite{Wei:2022zuf}, a CP violation observable, which is called the direct-CPA subtracted FB-CPA was introduced, which is defined as
\begin{equation*}
 \tilde{A}^{FB}_{CP}=\frac{\int_0^1 \left(\left\langle\left|\mathcal{M}\right|^2\right\rangle-\left\langle\left|\overline{\mathcal{M}}\right|^2\right\rangle \right)d\cos\theta-\int_{-1}^0\left(\left\langle\left|\mathcal{M}\right|^2\right\rangle-\left\langle\left|\overline{\mathcal{M}}\right|^2\right\rangle \right) d\cos\theta} {\int_{-1}^1\left(\left\langle \left|\mathcal{M}\right|^2\right\rangle+\left\langle\left|\overline{\mathcal{M}}\right|^2\right\rangle\right) d\cos\theta}.
\end{equation*}
When only $S$- and $P$- wave amplitudes involved, it has a simple form:
\begin{equation}
  \tilde{A}^{FB}_{CP}= \frac{\Re(\langle a_Pa_S^\ast\rangle)-\Re(\langle \overline{a_P}\overline{a_S}^\ast\rangle)}{\left\langle\left| a_P\right|^2\right\rangle/3+\left\langle\left| a_S\right|^2\right\rangle+\left\langle\left| \overline{a_P}\right|^2\right\rangle/3+\left\langle\left| \overline{a_S}\right|^2\right\rangle}.
\end{equation}
}
\begin{equation}
\begin{aligned}
A^{FB}_{CP}&=\frac{1}{2}\left(A^{FB}-\overline{A^{FB}}\right),
\label{eq3}
\end{aligned}
\end{equation}
where $\overline{A^{FB}}$ represents the FBA of the $CP$ conjugate process.

\section{\label{sec:DecayAmp} The decay amplitude of ${\overline{B}}^{0}\rightarrow K^{-}\pi^{+}\pi^{0}$}

When the invariant mass of the ${K^-\pi^+}$ system is around the masses of  $\overline{K}^{*}_{0}(700)$ and $\overline{K}^{*}(892)^{0}$, the decay will be dominated by the cascade decays $\overline{B}^{0}\rightarrow \overline{K}^{*}_{0}(700) (\to K^{-}\pi^{+})\pi^{0}$ and $\overline{B}^{0}\rightarrow \overline{K}^{*}(892)^{0} (\to K^{-}\pi^{+})\pi^{0}$.
The decay amplitudes can then be expressed as
\begin{equation}
\mathcal{M}_{{{\overline{B}}^{0}\rightarrow K^{-}\pi^{+}\pi^{0}}}=\mathcal{A}_{{K}_{0}^\ast}+\mathcal{A}_{{K}^{\ast 0}}e^{i\delta}\cos\theta,
\label{eq7}
\end{equation}
where $\mathcal{A}_{K^{*}_{0}}$ and $\mathcal{A}_{K^{\ast 0}}$ are the decay amplitudes for the two aforementioned cascade processes, and $\delta$ is the relative phase between them.
We can further isolate the Breit-Wigner factor, so that the amplitudes $\mathcal{A}_{{K}^{\ast}_{0}}$ and $\mathcal{A}_{{K}^{\ast 0}}$ can be re-expressed as:
\begin{equation}
  \mathcal{A}_X=\frac{\tilde{\mathcal{A}}_X}{s-m_{X}^2+im_X\Gamma_X},
\end{equation}
for $X={{K}^{\ast}_{0}}$, and ${{K}^{\ast 0}}$, where $s$ is the invariant mass squared of the $K^-\pi^+$ system, and
	\begin{equation}
		\begin{aligned}
			\tilde{\mathcal{A}}_{K^{\ast 0}}&={\sqrt{2}m_{K^{\ast 0}}g_{K^{\ast 0}K\pi}\left(\frac{s_{\pi^{0}\pi^{-}, \text{max}}-s_{\pi^{0}\pi^{-}, \text{min}}}{2}\right)} 
			\cdot\Bigg\{V_{ub}V^{*}_{us}a_{2}f_{\pi}A_{0}^{B\to K^{\ast 0}}\\
			&+V_{tb}V^{*}_{ts}\left[\left(a_{4}-\frac{1}{2}a_{10}\right)f_{K^{*}(892)^{0}}F_{1}^{B\to\pi} +\frac{3}{2}\left(a_{7}-a_{9}\right)f_{\pi}A_{0}^{B\to K^{\ast 0}}\right]\Bigg\},
			\label{eq4}
		\end{aligned}
	\end{equation}
	\begin{equation}
		\begin{aligned}
			\tilde{\mathcal{A}}_{K^{*}_{0}}={\sqrt{2}g_{K^{*}_{0}K\pi}} \cdot\Bigg( & -V_{ub}V^{*}_{us}  a_{2}(m^{2}_{B}-m^{2}_{K^\ast_0})f_{\pi}F_{0}^{B\to K^\ast_0}\\
			+ & V_{tb}V^{*}_{ts}\Bigg\{ 
			\left[a_{4}-\frac{1}{2}a_{10}-\frac{2m_{K^\ast_0}^2}{m_bm_s}\left(a_6-\frac{1}{2}a_8\right)\right](m^{2}_{B}-m^{2}_{\pi})f_{K^{*}_{0}}F_{0}^{B\to \pi} \\
			&  -\frac{3}{2}(a_{7}-a_9)(m^{2}_{B}-m^{2}_{{K}^\ast_0})f_{\pi}F_{0}^{B\to K^\ast_0}\Bigg\}\Bigg),
			\label{eq5}
		\end{aligned}
	\end{equation}
where $F_{0}^{B\to K^{\ast 0}}$, $A_{0}^{B\to K^\ast}$, $F_{0}^{B\to K^\ast_0}$, $F_0^{B\to\pi}$ and $F_{1}^{B\to \pi}$ are the form factors for the corresponding transitions $\overline{B}^0\to \overline{K}^\ast(892)^0$, $\overline{B}^0\to \overline{K}^\ast_0(700)$, and $\overline{B}^0\to \pi^0$,
$f_{K^{\ast 0}}$, $f_{K^{*}_{0}}$, $f_\pi$ are the decay constants, $s_{\pi^{0}\pi^{-},\text{max}}$ and $s_{\pi^{0}\pi^{-},\text{min}}$ are the maximum and minimum values of $s_{\pi^{0}\pi^{-}}$,
$g_{K^{*}_{0} K\pi}$ and $g_{K^{\ast 0} K\pi}$ are the strong coupling constants.
All the $a_{i}$'s are built up from the Wilson coefficients $c_{i}$'s,
and take the form $a_{i}=c_{i}+c_{i+1}/N_{c}^{\text{eff}}$ for odd $i$ and $a_{i}=c_{i}+c_{i-1}/N_{c}^{\text{eff}}$ for even $i$, where $N_{c}^{\text{eff}}$ is the effective color number, whose diviation from the color number $N_c=3$ mearsures the non-factorizable contribution.
The strong coupling constants $g_{K^{\ast 0} K\pi}$ and
$g_{K^{*}_{0} K\pi}$ are determined from the measured
partial decay widths through the relations \cite{Chua:2008zza}:
\begin{equation}
\begin{aligned}
 \Gamma_{K^{*}_{0} K\pi}={\frac{p_c}{8\pi m_{K^*_0}^2}}g_{K^{*}_{0} K\pi}^2,\qquad
 \Gamma_{K^{* 0} K\pi}=\frac{2}{3}\frac{p_c^3}{4\pi m_{K^{0*}}^2}g_{K^{*0} K\pi}^2,
\end{aligned}
\end{equation}
where $\Gamma_{K^{*}_{0} K\pi}$ and $\Gamma_{K^{*0} K\pi}$ are the decay widths for $K^{*}_{0}(700)\to K\pi$ and ${K^{*}(892)^{0}\to K\pi}$, respectively, and $p_c$ is the momentum of $K$ or $\pi$ in the rest frame of $K^{*}_{0}(700)$ or $K^{*}(892)_{0}$.

\section{\label{sec:NumDis} Numerical results and discussion}

\begin{table}[H]
\centering
\caption{\label{InputParameters}Input parameters used in this paper.}
\begin{tabular}{|*{8}{l|}}
\hline
Parameters&Input data & References\\
\hline
Wilson coefficients
&$c_1=1.1502,~c_{2}=-0.3125,~c_3=0.0174$& \cite{Deshpande:1994pw}\\
&$c_4=-0.0373,~c_5=0.0104,~c_6=-0.0459$&\\
&$c_7=-1.050\times10^{-5},~c_8=3.839\times10^{-4},~c_9=-0.0101$&\\
&$c_{10}=1.959\times10^{-3}$&\\
\hline
Form factors&$F^{B\rightarrow K^{*}_{0}(700)}_{0}=0.27 ,~F^{B\rightarrow\pi}_{0}=0.25,$&\cite{Cheng:2005nb,Qi:2018syl,Li:2008tk}\\
&$F_1^{B\to \pi}=0.25$, $A^{B\rightarrow K^{*}(892)^{0}}_{0}=0.374$&\\
\hline
Decay constants(in GeV)&$f_{\pi}={0.13}$, $f_{K^{*}(892)^{0}}=0.217,$ &\cite{ParticleDataGroup:2024cfk,Ball:2004rg,Narison:2008nj}\\
&$f_{K^{*}_{0}(700)}=0.064$& \\
\hline
masses, decay widths, and branching ratios & $m_{K^\ast_0(700)}=845$ MeV, $\Gamma_{K^\ast_0(700)}=468$ MeV,&\\
& $m_{K^\ast(892)^0}=895$ MeV, $\Gamma_{K^\ast(892)^0}=47$ MeV&\cite{ParticleDataGroup:2024cfk} \\ 
& $BR_{K^\ast_0(700) K\pi}=100\%$, $BR_{K^\ast(892)^0 K\pi}=100\%$& \\
\hline
\end{tabular}
\end{table}

\begin{figure}[H]
	\centering
	\subfloat[$\delta=0$]{
		\includegraphics[width=0.48\linewidth]{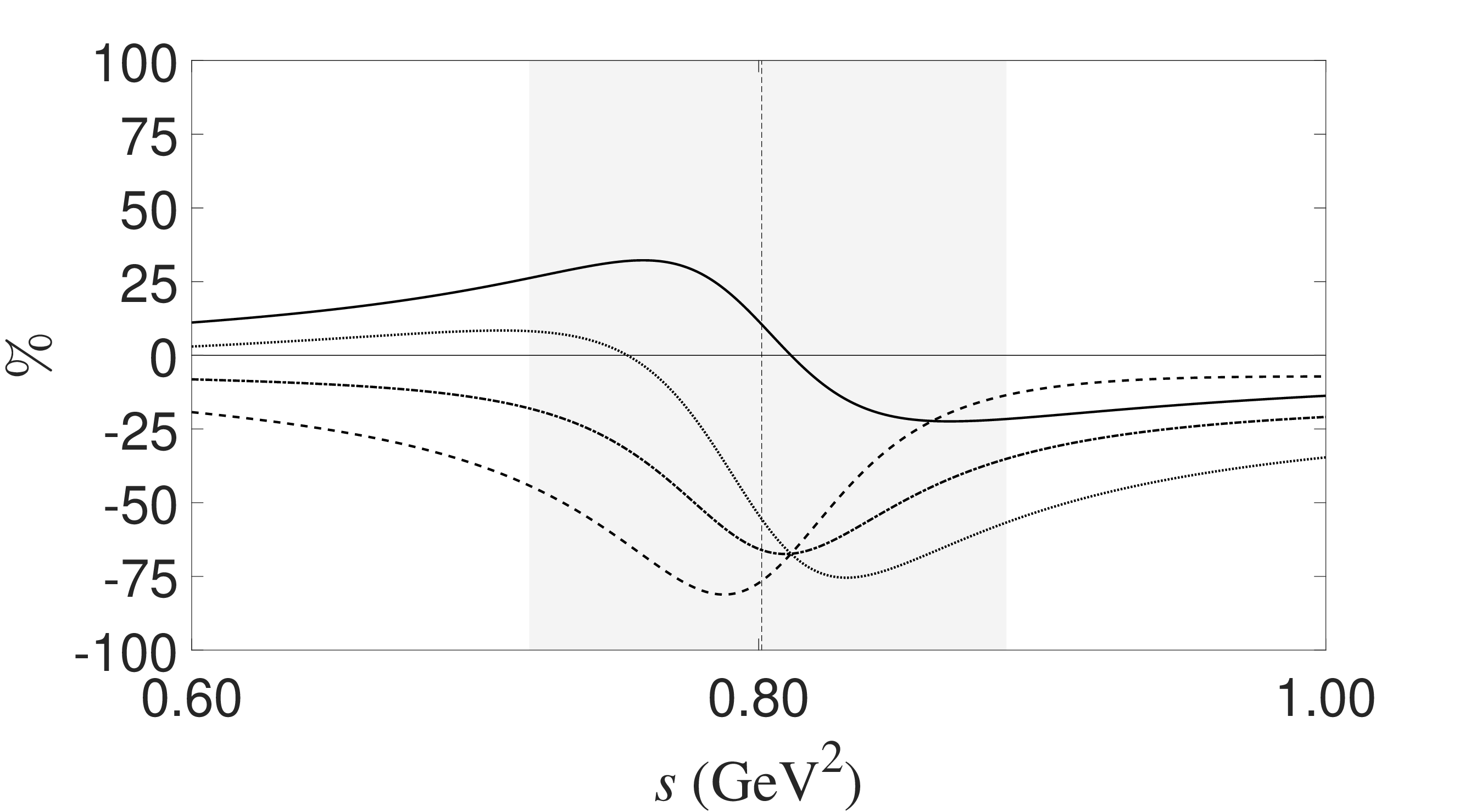}
	}\hfill
	\subfloat[$\delta=\frac{\pi}{2}$]{
		\includegraphics[width=0.48\linewidth]{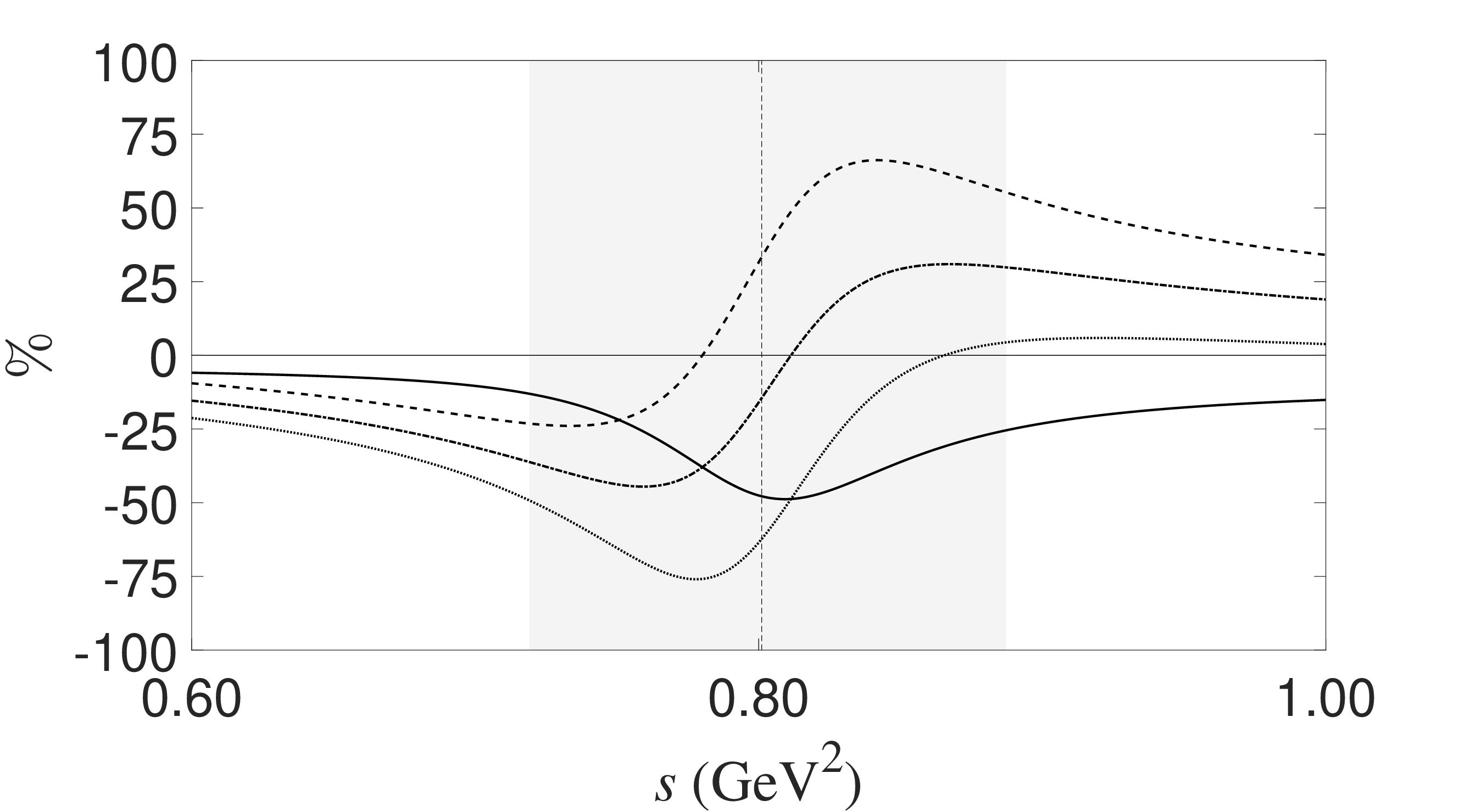}
	}\\
	\subfloat[$\delta={\pi}$]{
		\includegraphics[width=0.48\linewidth]{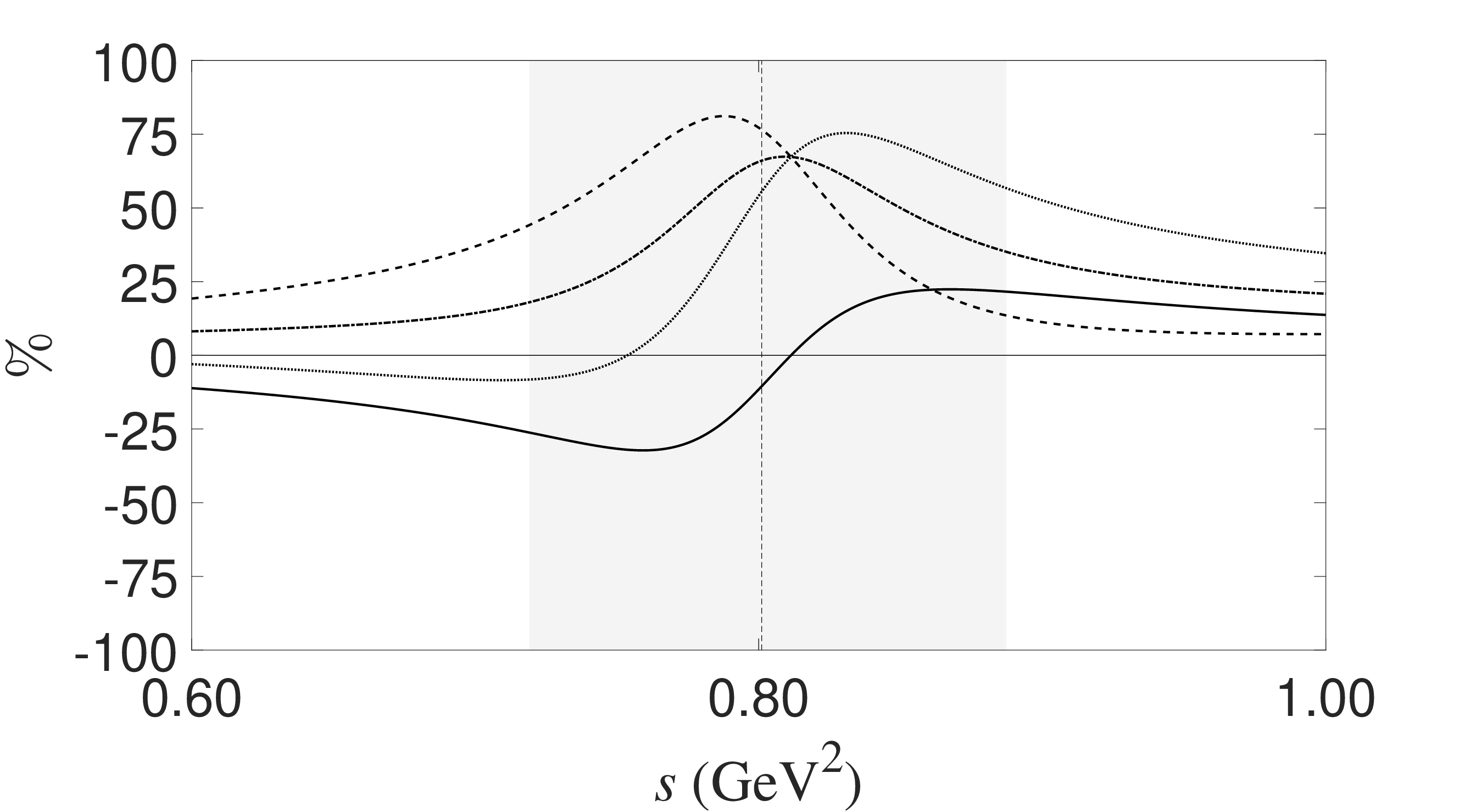}
	}\hfill
	\subfloat[$\delta=\frac{3\pi}{2}$]{
		\includegraphics[width=0.48\linewidth]{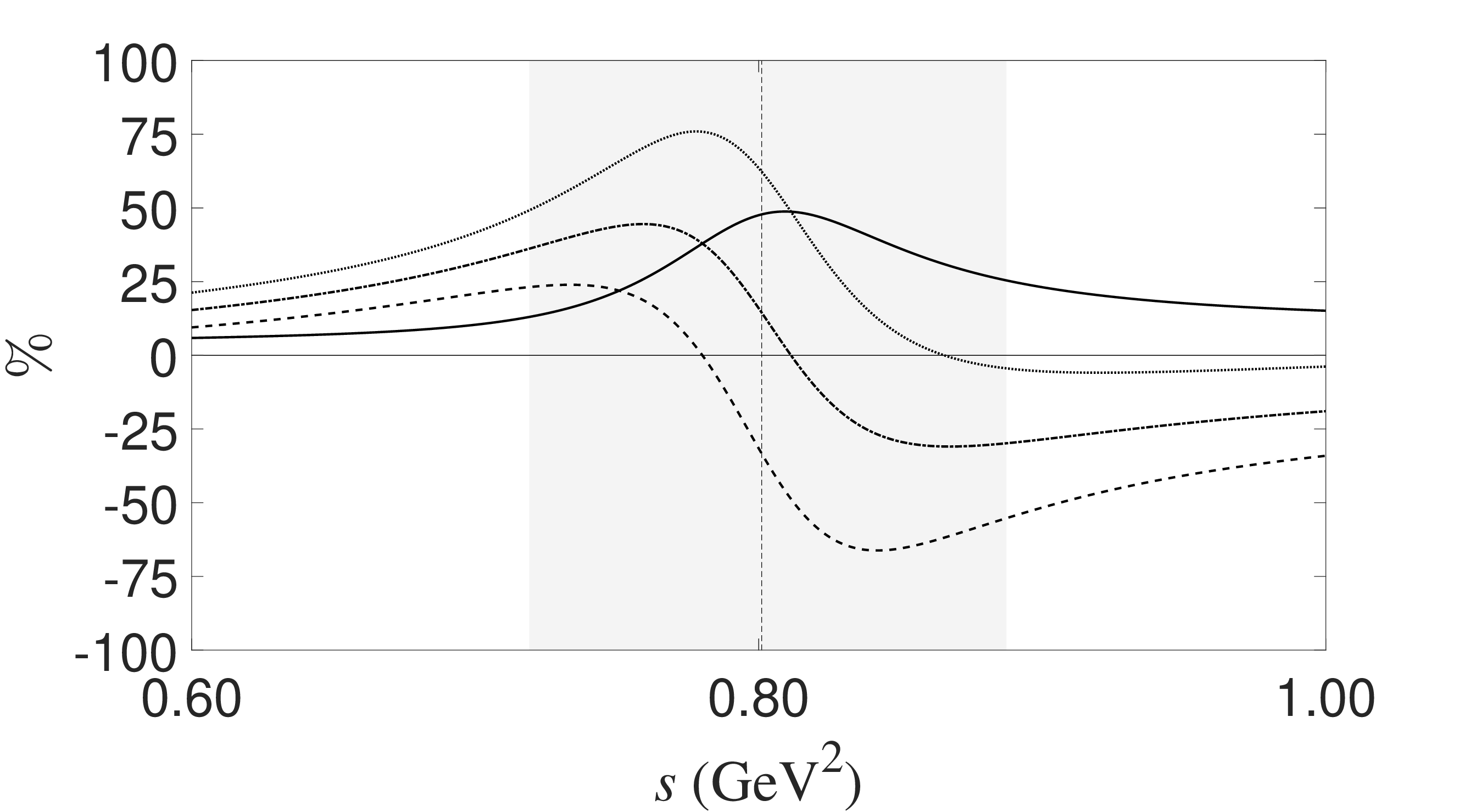}
	}\\
	\caption{The $s$-dependence of $A^{FB}$, $\overline{A^{FB}}$, $A^{FB}_{CP}$, and $A^{FB}_{\text{ave}}$ of the decay channel
		$B^{0}\rightarrow K^{+}\pi^{-}\pi^{0}$ for $\delta=0$, $\pi/2$, $\pi$, and $3\pi/2$, respectively, and for $N_{c}^{\text{eff}}=1$.
		The range of $s$ is taken from 0.4 GeV/$c^{2}$ to 1.2 GeV/$c^{2}$. 
		The dotted, dashed, dash-dotted and solid lines represent  $A^{FB}$, $\overline{A^{FB}}$, $A^{FB}_{\text{ave}}$, and $A^{FB}_{CP}$, respectively. 
		The shadowed region indicate the location of the vector resonances $K^{*}(892)^{0}$ (
		$(m_{K^{*}(892)^{0}}-\Gamma_{K^{*}(892)^{0}})^{2}<s<(m_{K^{*}(892)^{0}}+\Gamma_{K^{*}(892)^{0}})^{2}$).}
	\label{FIG.Nc1}
\end{figure}

\begin{figure}[H]
	\centering
	\subfloat[$\delta=0$]{
		\includegraphics[width=0.48\linewidth]{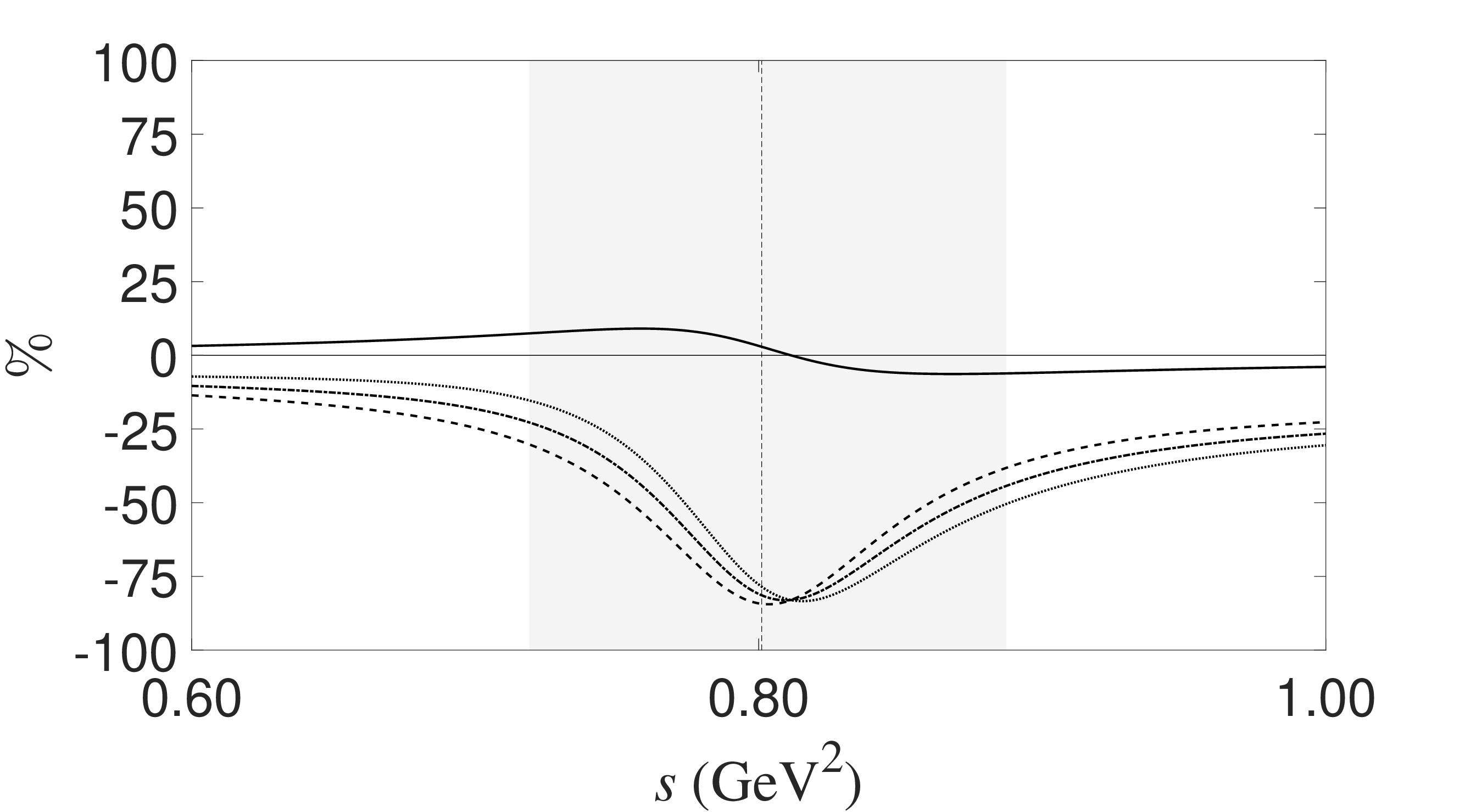}
	}\hfill
	\subfloat[$\delta=\frac{\pi}{2}$]{
		\includegraphics[width=0.48\linewidth]{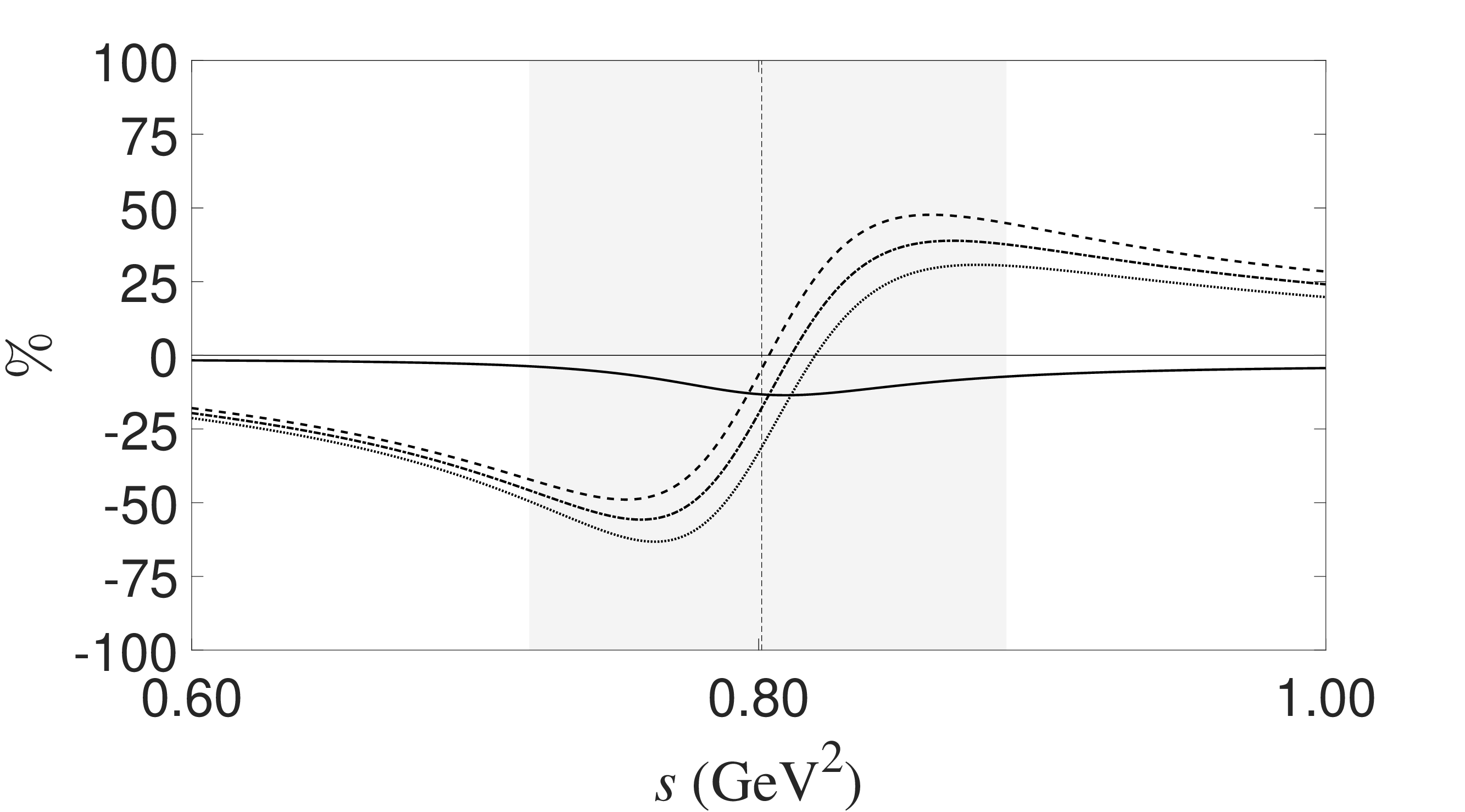}
	}\\
	\subfloat[$\delta={\pi}$]{
		\includegraphics[width=0.48\linewidth]{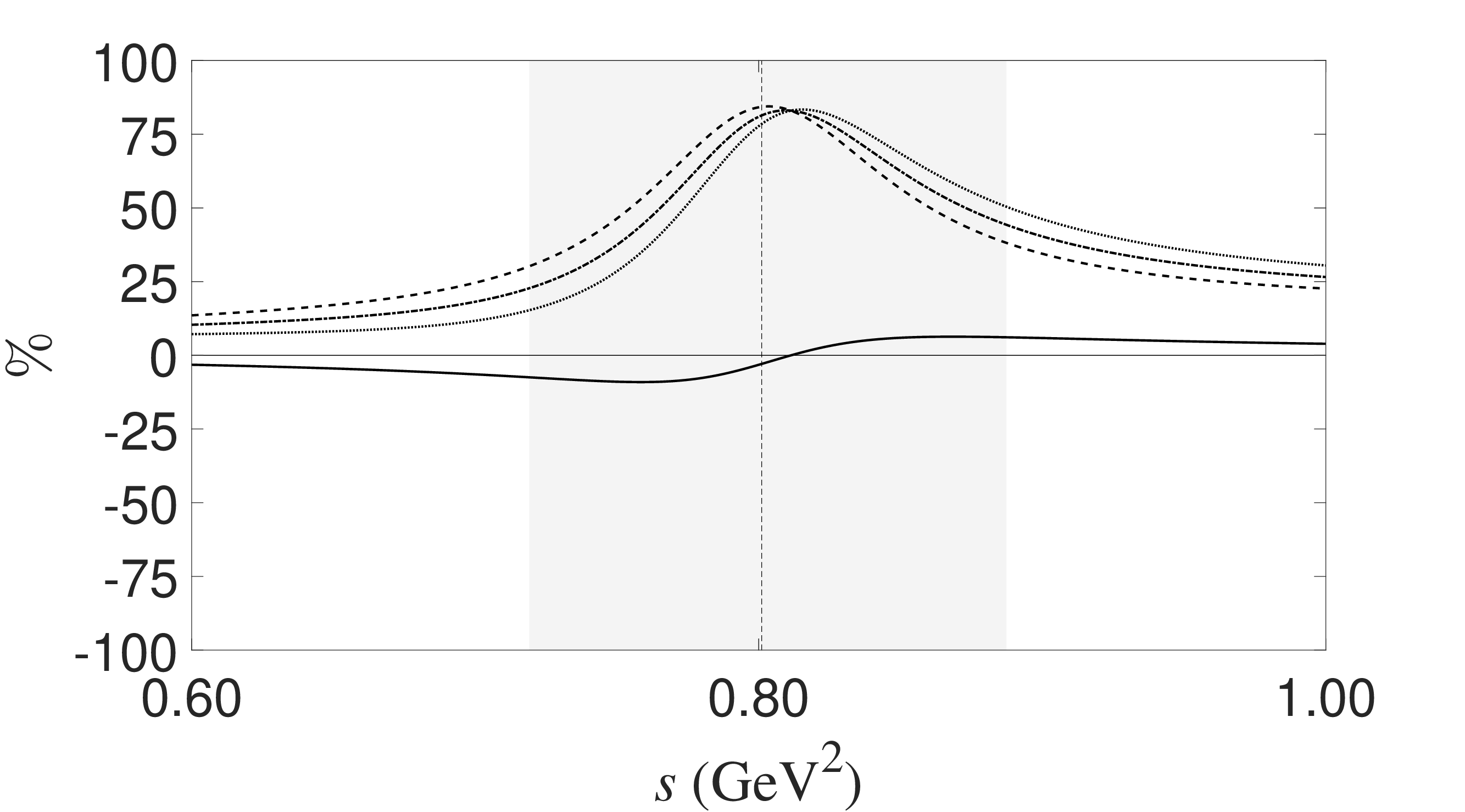}
	}\hfill
	\subfloat[$\delta=\frac{3\pi}{2}$]{
		\includegraphics[width=0.48\linewidth]{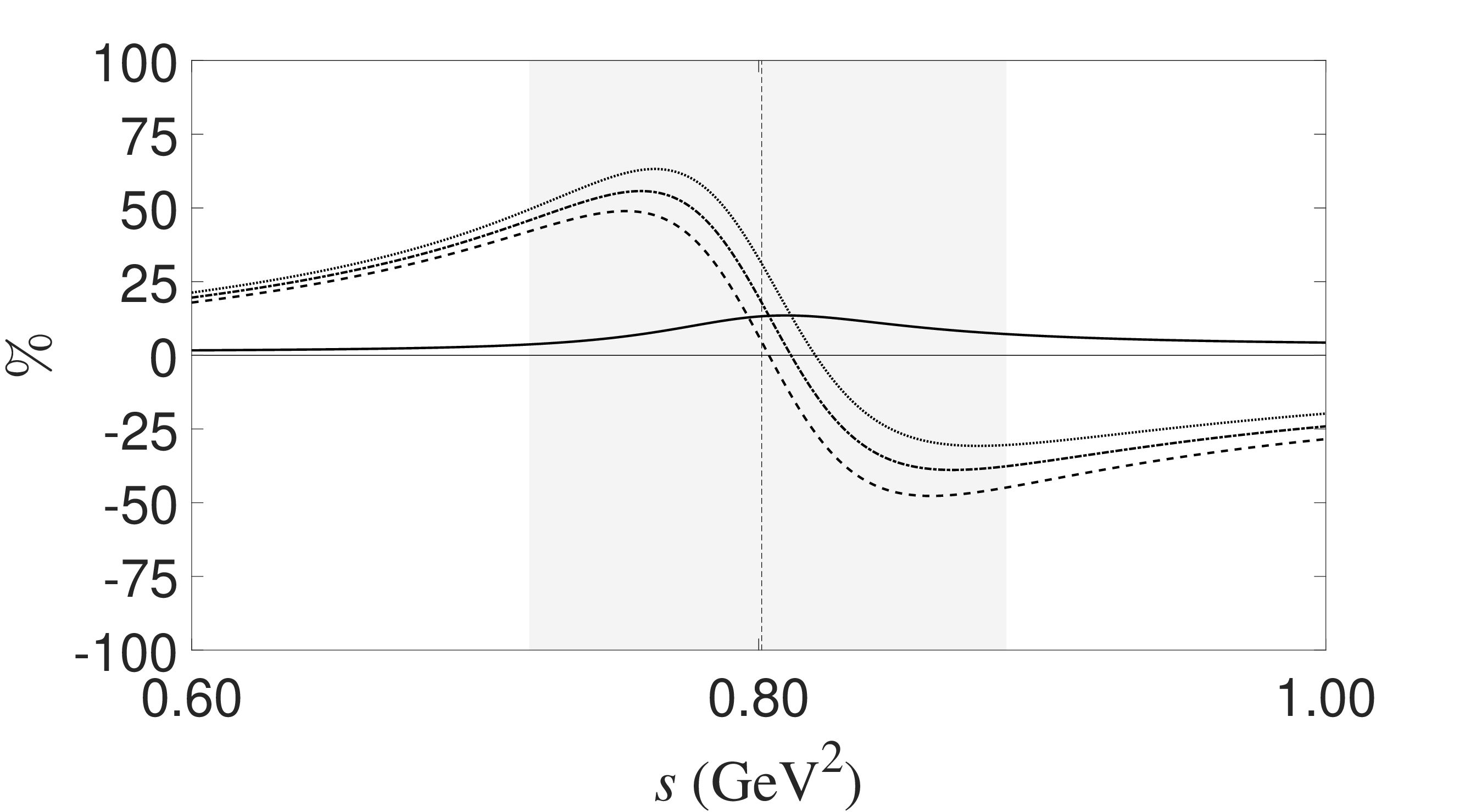}
	}\\
	\caption{The same as Fig. \ref{FIG.Nc1} but for $N_{c}^{\text{eff}}=2$.}
	\label{FIG.Nc2}
\end{figure}

\begin{figure}[H]
	\centering
	\subfloat[$\delta=0$]{
		\includegraphics[width=0.48\linewidth]{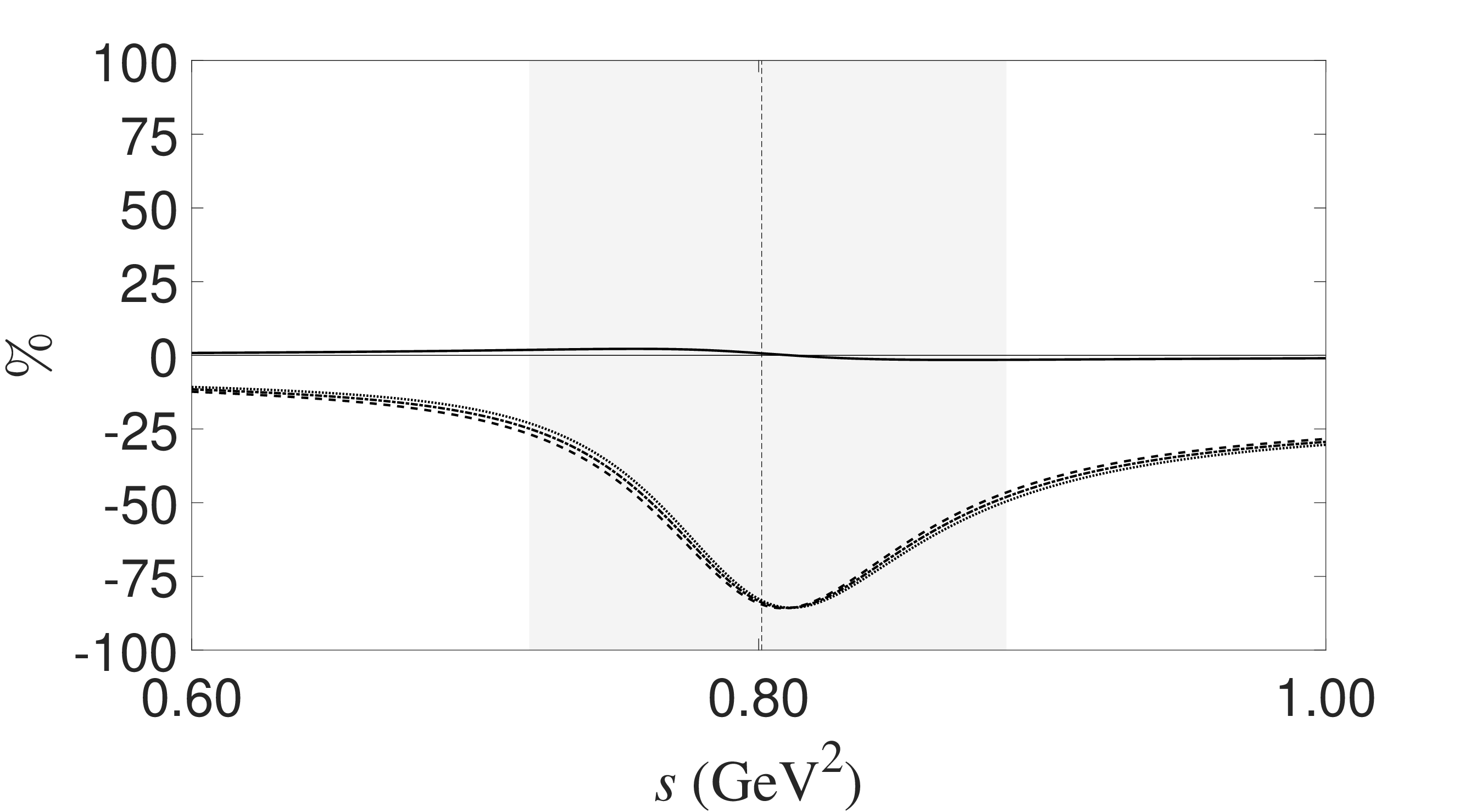}
	}\hfill
	\subfloat[$\delta=\frac{\pi}{2}$]{
		\includegraphics[width=0.48\linewidth]{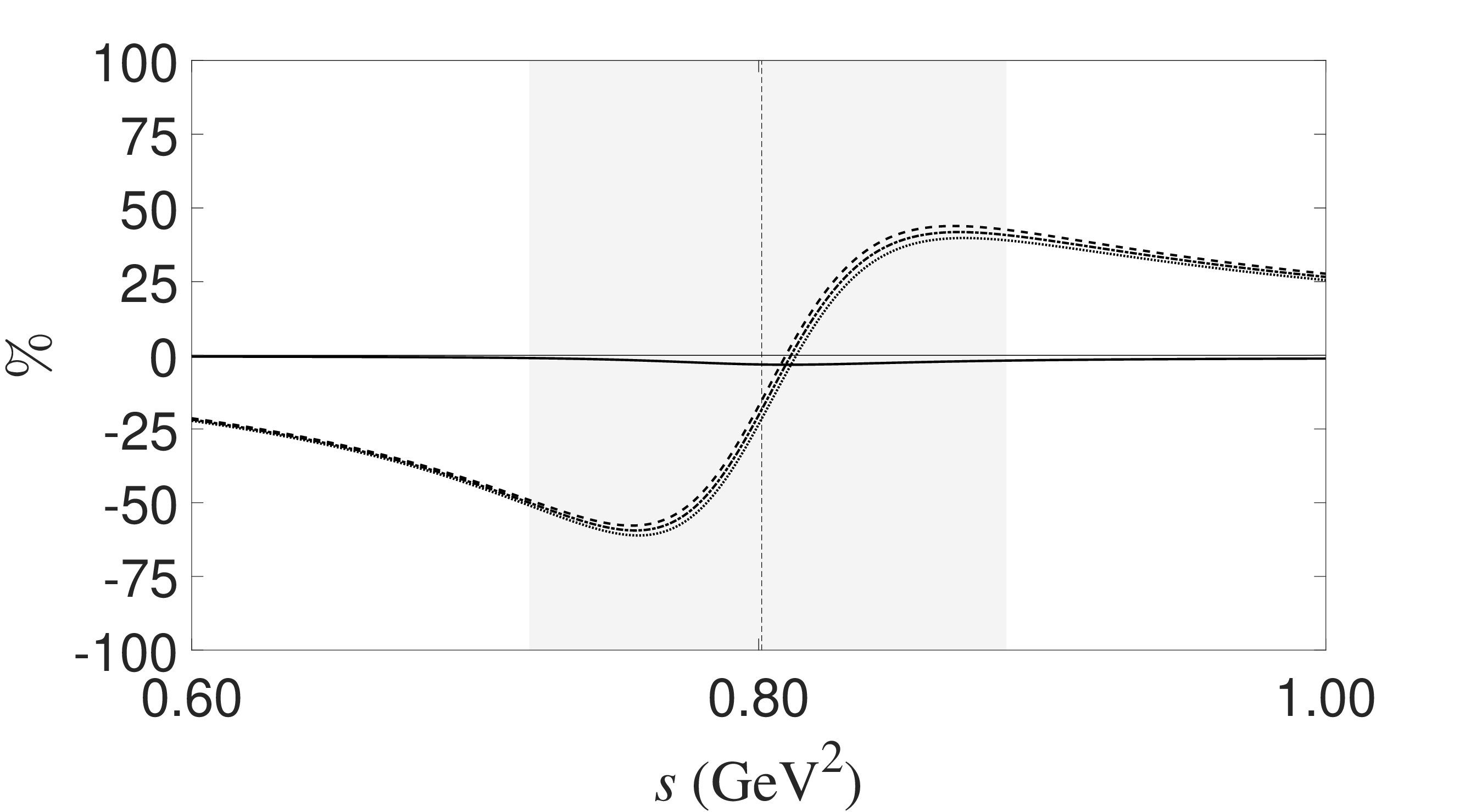}
	}\\
	\subfloat[$\delta={\pi}$]{
		\includegraphics[width=0.48\linewidth]{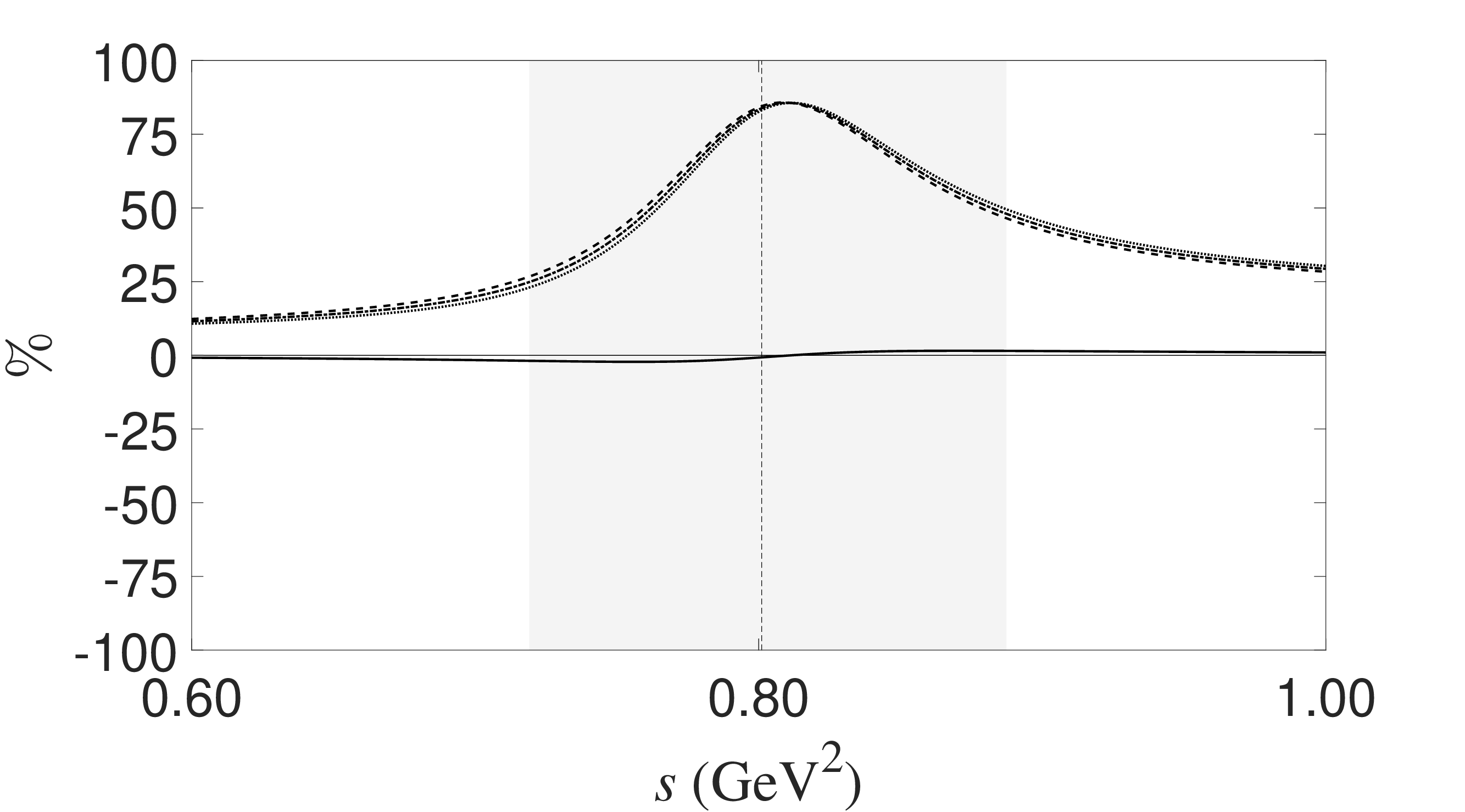}
	}\hfill
	\subfloat[$\delta=\frac{3\pi}{2}$]{
		\includegraphics[width=0.48\linewidth]{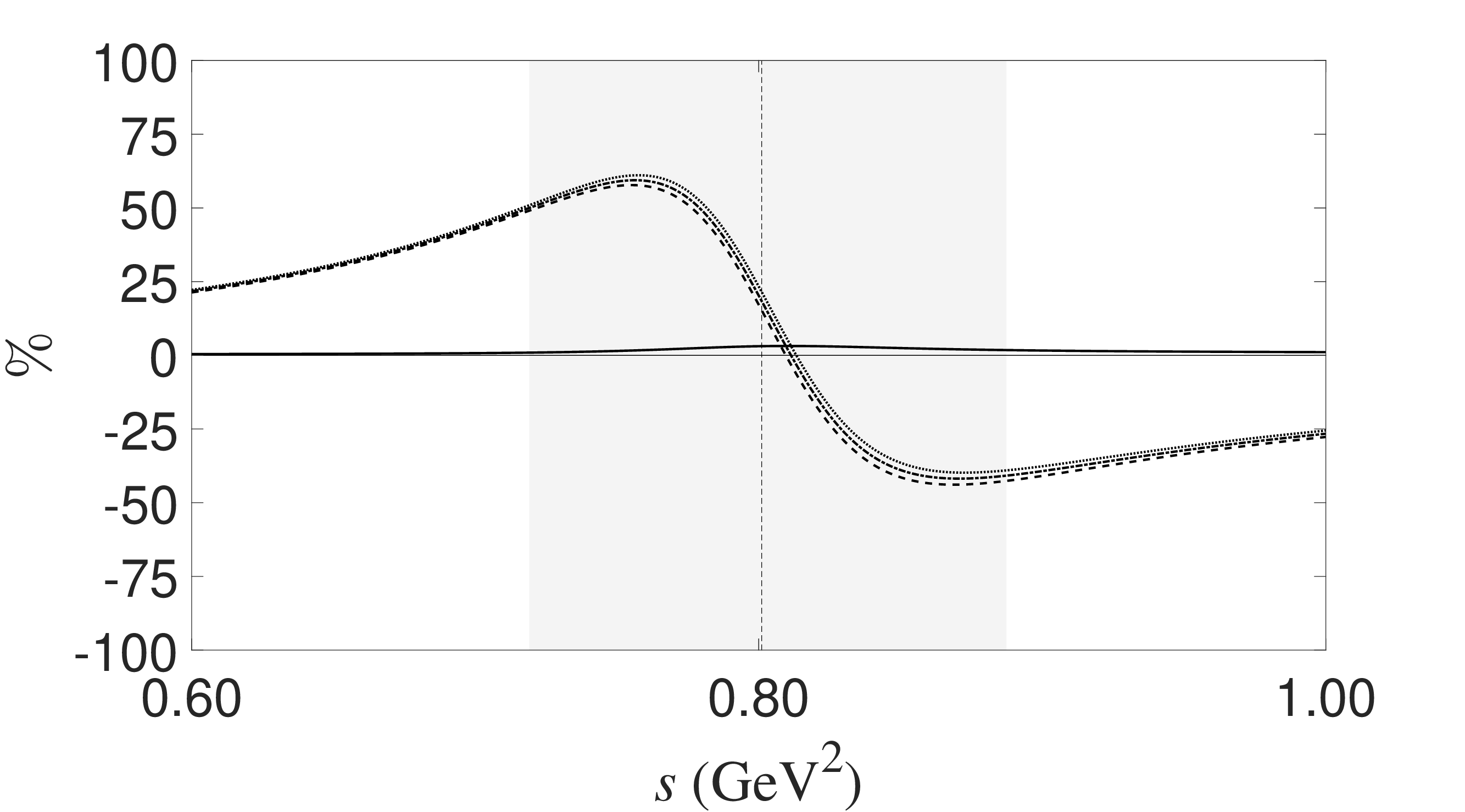}
	}\\
	\caption{The same as Figs. \ref{FIG.Nc1} and \ref{FIG.Nc2} but for $N_{c}^{\text{eff}}=3$.}
	\label{FIG.Nc3}
\end{figure}

In Figs. \ref{FIG.Nc1}, \ref{FIG.Nc2}, and \ref{FIG.Nc3} we present the AFBs of $B^0\to K^+\pi^-\pi^0$ and its $CP$-conjugate process $A^{FB}$, and $\overline{A^{FB}}$, and the corresponding FB-CPAs $A^{FB}_{CP}$, for $N_c^{\text{eff}}=1$, 2, and 3, respectively, for various values of the strong phase $\delta$, where the inputs of various parameters are listed in Table. \ref{InputParameters}.
One can see from all these figures that $A^{FB}$ and $\overline{A^{FB}}$ tend to take large values when $s$ is around the mass-squared of $\overline{K}^\ast(892)^0$ and $\overline{K}^\ast_0(700)$, implying that the interference effect of $\overline{K}^\ast(892)^0$ and $\overline{K}^\ast_0(700)$ induces large FBAs.
Note that this behaviour is universal to different values of $N_c^{\text{eff}}$. 
On the other hand, the CP violation observables FB-CPAs are much more sensitive to the values of $N_c^{\text{eff}}$.
By comparing these three figures one can immediately see that a non-zero FB-CPAs lies in the regions of $\overline{K}^{\ast}(892)$ and $\overline{K}^\ast_0(700)$ for $N_c^{\text{eff}}=1$ and 2, as can be seen from Figs. \ref{FIG.Nc1} and \ref{FIG.Nc2}, while the FB-CPAs is negligibly small for $N_c^{\text{eff}}=3$ from Fig. \ref{FIG.Nc3}, indicating an important impact of non-factorizable contributions, which are characterized by $N_c^{\text{eff}}$, to the FB-CPAs.

Figs. \ref{FIG.Nc1}, \ref{FIG.Nc2}, and \ref{FIG.Nc3} also show an important fact that the behaviour of FB-CPA, as a function of $s$,  strongly depends on the value $\delta$.
One important property is that, for $\sqrt{s}$ bellow and above the mass of the vector resonance $\overline{K}^\ast(892)^0$, when $\delta$ takes values around $0$ or $\pi$, the FB-CPA tends to take opposite values; while when $\delta$ takes values around $\pi/2$ or $3\pi/2$, the FB-CPA does not change the sign. 
This behaviour can be understood as follows.

For simplicity, in what follows, we will denote  ${\overline{K}^\ast_0(700)}$ and $\overline{K}^\ast(892)^0$ as ${S}$ and ${P}$, respectively, to reflect that the angular momentum of the ${\overline{K}^\ast_0(700)} \pi^0$ and ${\overline{K}^\ast(892)^0}\pi^0$ system are $l=0$ ($S$) and $l=1$ ($P$), respectively.
By isolating the Breit-Wigner factors, the interference term in $A^{FB}$ can be expressed as
\begin{equation}\label{eq:SPinterfer}
\Re( \mathcal{A}_{S}^\ast \mathcal{A}_{P} e^{i\delta})=\lambda\left\{\left[1+\frac{(s-m_P^2)(s-m_S^2)}{m_P\Gamma_Pm_S\Gamma_S}\right]\Re (\tilde{\mathcal{A}}_{S}^\ast\tilde{\mathcal{A}}_{P} e^{i\delta})
+\left[-\frac{s-m_P^2}{m_P\Gamma_P}+\frac{s-m_S^2}{m_S\Gamma_S}\right]\Im (\tilde{\mathcal{A}}_{S}^\ast\tilde{\mathcal{A}}_{P} e^{i\delta})\right\}.
\end{equation}
where
\begin{equation}
\lambda=\frac{m_P\Gamma_Pm_S\Gamma_S}{\left[\left(s-m_{P}^2\right)^2+ \left(m_P\Gamma_P\right)^2\right]\left[\left(s-m_{S}^2\right)^2+\left(m_S\Gamma_S\right)^2\right]}. 
\end{equation}
Note that $\Gamma_S$ is considerably larger than $\Gamma_P$. Meanwhile, since we focus mainly on the region when $s$ is around the mass-squared of $S$ and $P$, we have $\Gamma_s \gg \sqrt{|s-m_S^2|}$ and $\Gamma_s \gg \sqrt{|s-m_P^2|}$.
This means that we can safely approximate Eq. (\ref{eq:SPinterfer}) as
\begin{equation}
\Re( \mathcal{A}_{S}^\ast \mathcal{A}_{P} e^{i\delta})=\lambda\left[\Re (\tilde{\mathcal{A}}_{S}^\ast\tilde{\mathcal{A}}_{P} e^{i\delta})
-\frac{s-m_P^2}{m_P\Gamma_P}\Im (\tilde{\mathcal{A}}_{S}^\ast\tilde{\mathcal{A}}_{P} e^{i\delta})+\mathcal{O}(\epsilon)\right],
\end{equation}
where we have collectively denote the three small quantities, $\Gamma_P/\Gamma_S$, $\sqrt{\left|s-m_S^2\right|}/\Gamma_S$, and $\sqrt{\left|s-m_P^2\right|}/\Gamma_S$, as $\epsilon$.

Up to the order $\mathcal{O}(\epsilon)$, $\Re( \mathcal{A}_{S}^\ast \mathcal{A}_{P} e^{i\delta})$ can be further splitted into two parts
\begin{equation}
\Re( \mathcal{A}_{S}^\ast \mathcal{A}_{P} e^{i\delta})=\lambda\left(\Delta+\Sigma\right),
\end{equation}
where
\begin{equation}\label{eq:Delta}
\Delta=-\left(\sin\delta
+\frac{s-m_P^2}{m_P\Gamma_P}\cos\delta\right)\Im (\tilde{\mathcal{A}}_{S}^\ast\tilde{\mathcal{A}}_{P}),
\end{equation}
and
\begin{equation}\label{eq:Sigma}
\Sigma= \left(\cos\delta
-\frac{s-m_P^2}{m_P\Gamma_P}\sin\delta\right) \Re (\tilde{\mathcal{A}}_{S}^\ast\tilde{\mathcal{A}}_{P}).
\end{equation}
To explain the motivation for this splitting, notice that if $\delta$ is the only dominate strong phase, one will have $\tilde{\mathcal{A}}_{S}^\ast\approx\tilde{\mathcal{A}}_{S}^{CP}$, and $\tilde{\mathcal{A}}_{P}^\ast\approx\tilde{\mathcal{A}}_{P}^{CP}$, where $\tilde{\mathcal{A}}_{S}^{CP}$ and $\tilde{\mathcal{A}}_{S}^{CP}$ are the corresponding amplitudes for the $CP$-conjugate processes. 
As a result, one has
$\Im (\tilde{\mathcal{A}}_{S}^\ast\tilde{\mathcal{A}}_{P})\approx-\Im (\tilde{\mathcal{A}}_{S}^{CP\ast}\tilde{\mathcal{A}}_{P}^{CP})$
and 
$\Re (\tilde{\mathcal{A}}_{S}^\ast\tilde{\mathcal{A}}_{P})\approx \Re (\tilde{\mathcal{A}}_{S}^{CP\ast}\tilde{\mathcal{A}}_{P}^{CP})$.
Consequently one has $\Delta\approx-\Delta^{CP}$ and $\Sigma\approx\Sigma^{CP}$, where $\Delta^{CP}$ ($\Sigma^{CP}$) is the same as $\Delta$ ($\Sigma$) except that the amplitudes are replaced by the CP-conjugate ones.
This means that if  $\delta$ is the only dominate strong phase, FB-CPA 
will be dominated by $\Delta$, so that
\begin{equation}
  A^{FB}_{CP} \sim \Re( \mathcal{A}_{S}^\ast \mathcal{A}_{P} e^{i\delta})-\Re( \mathcal{A}_{S}^{CP\ast} \mathcal{A}_{P}^{CP} e^{i\delta})\sim (\Delta-\Delta^{CP})+(\Sigma-\Sigma^{CP})\approx \Delta-\Delta^{CP}\approx 2\Delta.
\end{equation}

Now the different behaviours of FB-CPA in the four subfigures of Figs. \ref{FIG.Nc1}, \ref{FIG.Nc2}, and \ref{FIG.Nc3} can be understood. 
The two terms in $\Delta$ of Eq. (\ref{eq:Delta}) have quite different behaviour. 
The second term changes sign when $s$ passing through $m_P^2$ because of the presence of the factor $s-m_P^2$, while the first term does not.
When $\delta$ take values around $0$ or $\pi$, the second term dominates, so that the FB-CPA tends to change sign  when $s$ passing through $m_P^2$; when  $\delta$ take values around $\pi/2$ or $3\pi/2$, the first term dominates, so that the FB-CPA tends to not change sign throughout the interference region
\footnote{
The behaviour of changing of sign of $A^{FB}_{CP}$ as $s$ varing allows us to construct CPA observables corresponding the term $\Im (\tilde{\mathcal{A}}_{S}^\ast\tilde{\mathcal{A}}_{P} e^{i\delta})$ \cite{Qi:2024zau}. 
LHCb collabaration has searching for CPV in $D\to K K \pi$ \cite{LHCb:2013gqz,LHCb:2024rkp}. In the phase space where $\phi(1020)$ locates, they looked for CPV via an observable $A_{CP|S}$. 
This observable is the same in essence as that proposed in Ref. \cite{Qi:2024zau}.
For the current situation, the interference of vector and pseudo-scalar resonances may result in a change of sign of $A^{FB}_{CP}$ (as well as regional CPA) provided that the strong phase $\delta$ takes values around $0$ or $\pi$.
}.

Note that in Figs. \ref{FIG.Nc1}, \ref{FIG.Nc2}, and \ref{FIG.Nc3} we also present the $CP$-averaged FBA, which is defined as
\begin{equation}
\begin{aligned}
 A^{FB}_{\text{ave}}=\frac{\int_0^1\langle\left|\mathcal{M}\right|^2+\left|\overline{\mathcal{M}}\right|^2\rangle d\cos\theta-\int_{-1}^0\langle\left|\mathcal{M}\right|^2+\left|\overline{\mathcal{M}}\right|^2\rangle d\cos\theta} {\int_{-1}^1\langle \left|\mathcal{M}\right|^2+\left|\overline{\mathcal{M}}\right|^2\rangle d\cos\theta}.
\label{eq12}
\end{aligned}
\end{equation}
It can be easily shown that the $CP$-averaged FBA is dominated by $\Sigma$, $A^{FB}_{\text{ave}}\sim 2\Sigma$,
if $\delta$ is the only dominate strong phase.
As can be seen from Eq. (\ref{eq:Sigma}) that $A^{FB}_{\text{ave}}$ tends to change sign for $\delta$ taking values around $\pi/2$ or $3\pi/2$, while the sign of $A^{FB}_{\text{ave}}$ remains unchanged for $\delta$ taking values around $0$ or $\pi$, in contrast to the situation of $A^{FB}_{CP}$ 
\footnote{
It is also possible that $\tilde{\mathcal{A}}_{S}^\ast\tilde{\mathcal{A}}_{P}$ is almost real, i.e.,  $ \Re (\tilde{\mathcal{A}}_{S}^\ast\tilde{\mathcal{A}}_{P})\gg\Im (\tilde{\mathcal{A}}_{S}^\ast\tilde{\mathcal{A}}_{P})$. This happens either because of the weak phase is very small, such as the $D$ meson decay processes, or extra strong phases provide cancellation to result in a small imaginary part of $\tilde{\mathcal{A}}_{S}^\ast\tilde{\mathcal{A}}_{P}$. In this kind of situations, the behaviour of $A^{FB}$ and $\overline{A^{FB}}$ is dominated by $\Sigma$, which can change or not change of sign when $s$ passing through $m_P$, but $A_{CP}^{FB}$ remains small.}.

\begin{figure}[H]
	\centering
	\includegraphics[width=0.8\linewidth]{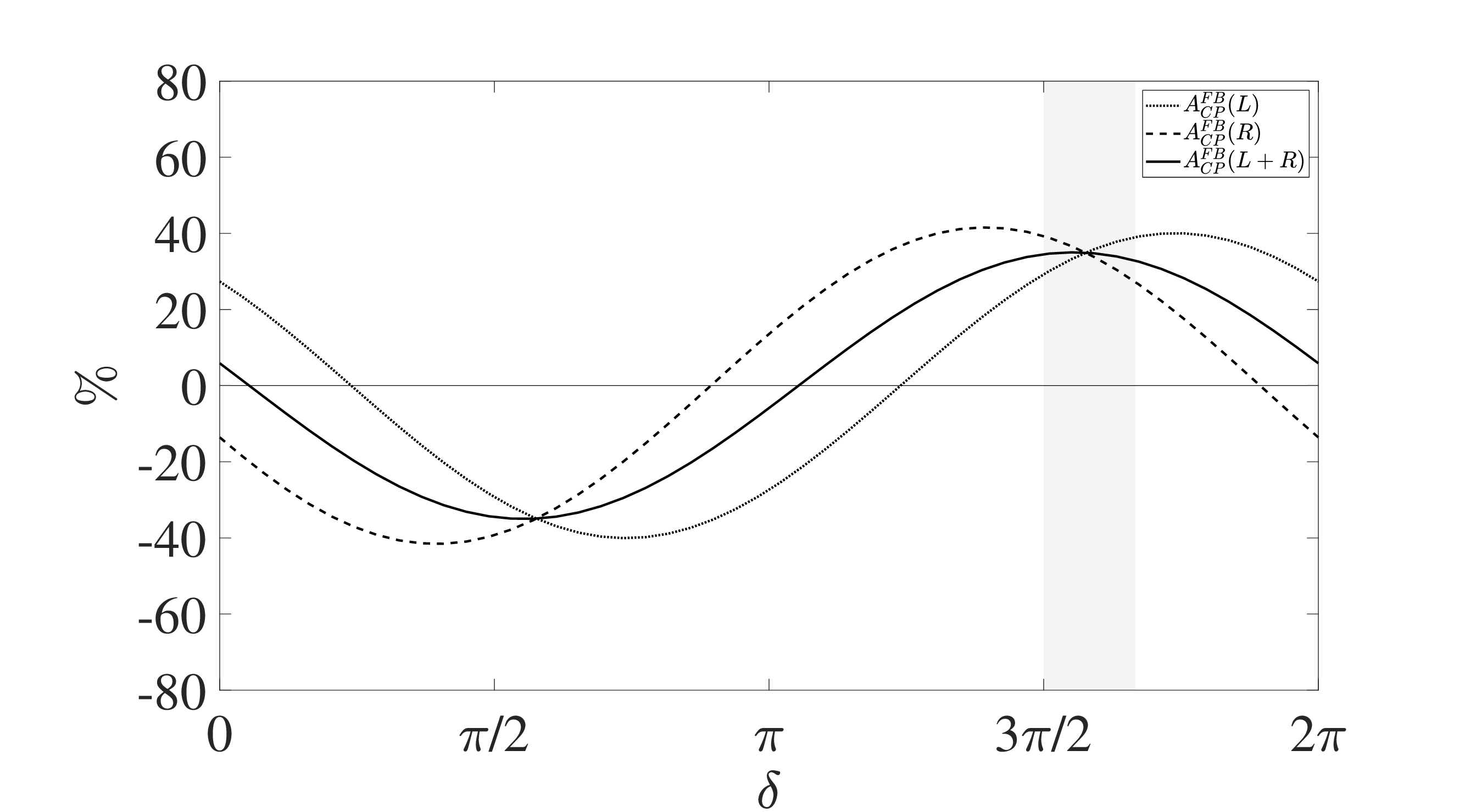}
	\caption{The FB-CPAs of three different regions in phase space, $L$ ($m_P-\Gamma_P<s<m_P$), $R$ ($m_P<s<m_P+ \Gamma_P$), and $L+R$ ($m_P-\Gamma_P<s<m_P+\Gamma_P$), which are respectively denoted as $A^{FB}_{CP}(L)$, $A^{FB}_{CP}(R)$, and $A^{FB}_{CP}(L+R)$ in the maintext and are represented by dotted, dashed, and solid lines, respectively, as functions of the strong phase $\delta$ for $N_c^{\text{eff}}=1$.
		The shadowed area indicates the range of $\delta$ preferred by the data in Ref. \cite{BaBar:2007hmp}.}
	\label{fig:AFBdelta}
\end{figure}

The aforementioned behaviour is also indicated in Fig. \ref{fig:AFBdelta}, in which we show the $\delta$-dependence of FB-CPAs for $N_c^{\text{eff}}=1$ in different phase-space regions: $m_P-\Gamma_P<s<m_P$, $m_P<s<m_P+ \Gamma_P$, and $m_P-\Gamma_P<s<m_P+\Gamma_P$, which will be denoted as region $L$, region $R$ and region $L+R$, respectively.
From Fig. \ref{fig:AFBdelta} one see that when $\delta$ takes values around 0 or $\pi$, the value of $A^{FB}_{CP}(L)$ is quite different from that of $A^{FB}_{CP}(R)$, and they tends to take opposite values.
While  when $\delta$ takes values around $\pi/2$ or $3\pi/2$, the difference between the values of $A^{FB}_{CP}(L)$ and $A^{FB}_{CP}(R)$ is small.

A Dalitz analysis of the decay $B^0 (\overline{B^0})\to K^\pm\pi^\mp \pi^0$ has been performed by the BaBar collaboration \cite{BaBar:2007hmp}.
The result of Fig. 7 in Ref. \cite{BaBar:2007hmp} indicates that  the $CP$-averaged FBA $A^{FB}_{\text{ave}}$ defined here tends to take positive(negative)  values  when $s$ is below (above) $m_P^2$.
By a closer comparison of the result of Fig. 7 in Ref. \cite{BaBar:2007hmp} with the $CP$-averaged FBA here, it can be shown that the strong phase $\delta$ is favoured to take values roughly in the range $[3\pi/2, 5\pi/3]$.
This means that the subfigures (d) in Figs. \ref{FIG.Nc1}, \ref{FIG.Nc2}, and \ref{FIG.Nc3} are more favourable by the data than the other three.
As can be seen from Fig. \ref{fig:AFBdelta}, for $\delta$ taking values in the range $[\pi/2, 2\pi/3]$, the corresponding FB-CPA in the region $L+R$ can takes value as large as about 35\% for $N_c^{\text{eff}}=1$, which are very likely accessible for Bell and Bell-II. 

In the above analysis for the interference effects of $\overline{K}^\ast(892)^0$ and $\overline{K}_0^\ast(700)$ in $\overline{B}^0\to K^-\pi^+ \pi^0$, we have used the condition that the strong phase $\delta$ is the only dominate strong phase.
In other words, there are no large physical strong phase within the amplitudes for each of the cascade decays,
which means that the direct CP asymmetries for the two-body decay processes $B^0\to K^\ast(892)^0\pi^0$ and $B^0\to K_0^\ast(700)\pi^0$ are negligibly small.
This is consistent with BaBar's experimental result \cite{BaBar:2007hmp,BaBar:2011vfx}.
This is true in the naive factorization approach for the weak decay processes, which is what we have adopted in this paper.
However, beyond the naive factorization, it is also possible that there are large strong phase within the amplitudes for the cascade decays.
In this kind of situations, non-negligible CPAs are expected in the aforementioned two-body decays, and
the above analysis will be contaminated.
It should be point out that whether this is the case can be examined by comparing FB-CPA and the corresponding direct-CPA-subtracted FB-CPA.
If there are obvious difference between them, that must means that at least one of the the direct CPAs for the aforementioned two two-body decays is relatively not small.
If relax the assumption of the dominance of the strong phase $\delta$, a combined analysis of the behaviour of FBA, CP-averaged FBA, FB-CPA, direct-CPA-subtracted FB-CPA, as well as the regional CPAs can also give us a more comprehensive understanding of the underlying dynamics of CP violation in the interference region. 

Our analysis here is in fact quite general, and can be applied to other three-body decay processes.
Take the well studied $B^\pm\to\pi^\pm\pi^+\pi^-$ channel as an example.
It has been shown experimentally that the interference of $\rho(770)^0$ and $f_0(500)$ has great impact to CPV in this decay channel.
The regional CPAs change sign when the invariant mass squared of the $\pi^+\pi^-$ system $s_{\pi^+\pi^-}$ passing through the mass squared of $\rho(770)^0$ \cite{LHCb:2019jta}.
This behaviour can be well explained similarly via a term just the same as that in Eq. (\ref{eq:SPinterfer}), since this kind of term can also contributes to regional CPAs.    
We strongly suggest our experimental colleagues to performed combined analysis of the aforementioned observables in three-body decays of heavy mesons in the interference region of intermediate resonances.

\section{\label{sec:SumCon} Summary and Conclusion}

In this paper, we present an analysis of CP violation induced by the interference effect of the intermediate resonances $\overline{K}^{*}(892)^{0}$ and $\overline{K}^{*}_{0}(700)$ in three-body decay $\overline{B}^{0}\rightarrow K^{-}\pi^{+}\pi^{0}$, based on the naive factorization approach for the weak two-body decay processes $\overline{B}^{0}\rightarrow \overline{K}^{*}(892)^{0}\pi^{0}$ and $\overline{B}^{0}\rightarrow \overline{K}^{*}_{0}(700)\pi^{0}$.
The interference between $\overline{K}^{*}(892)^{0}$ ($J^P=1^-$) and $\overline{K}^{*}_{0}(700)$ ($J^P=0^+$) can generate significant FBA.
This FBA further induces potentially measurable CP violation, with FB-CPA values reaching about 35\% for phenomenologically reasonable parameters.

According to the analysis here, the non-factorizabe contributions as we as the relative strong phase $\delta$ between the amplitudes corresponding to the cascade decays plays an important role in the behaviour of FBA and FB-CPA.
We present a general analysis of the correlation between the non-factorizabe contributions, the values of the strong phase $\delta$ and the behaviour for FBAs and FB-CPAs.
We discover a model-independent correlation between FBA and FB-CPA that persists across parameter variations.

The results of this paper establish a framework of the inclusion of the $\overline{K}^{*}(892)^{0}-\overline{K}^{*}_{0}(700)$ interference effect when studying CPV in multi-body decays, with both theoretical significance and immediate experimental relevance.
This framework explains similar CPV observations in $\Lambda_b\to p K^- \pi^+\pi^-$ and $\overline{B}^0 \to p \overline{p} K^-\pi^+$ decays.
The analysis can also apply to other multi-body decays of bottom and charmed hadrons.

\begin{acknowledgments}
This work was supported by National Natural Science Foundation of China under Grants Nos. 12475096, 12405115, 12192261, 
and Scientific Research Fund of Hunan Provincial Education Department under Grants No. 22A0319.
\end{acknowledgments}

\bibliography{zyj}

\end{document}